\documentclass[prd,aps,floatfix,twocolumn,showpacs,nofootinbib]{revtex4}

\usepackage[dvipdfm]{graphicx}
\usepackage{amsmath,amssymb}
\usepackage{mathrsfs}
\usepackage{natbib}
\usepackage{color}
\usepackage{bm}
\usepackage{txfonts}

\input{colordvi.tex}

\newcommand{\simgt}{\lower.5ex\hbox{$\; \buildrel > \over \sim \;$}}
\newcommand{\simlt}{\lower.5ex\hbox{$\; \buildrel < \over \sim \;$}}

\def\pd{{\rm d}}

\def\reff@jnl#1{{\rm#1\/}}
\def\aj{\reff@jnl{AJ}}                  
\def\araa{\reff@jnl{ARA\&A}}            
\def\apj{\reff@jnl{ApJ}}                
\def\apjl{\reff@jnl{ApJ}}               
\def\apjs{\reff@jnl{ApJS}}              
\def\ao{\reff@jnl{Appl.Optics}}         
\def\apss{\reff@jnl{Ap\&SS}}            
\def\aap{\reff@jnl{A\&A}}               
\def\aapr{\reff@jnl{A\&A~Rev.}}         
\def\aaps{\reff@jnl{A\&AS}}             
\def\azh{\reff@jnl{AZh}}                
\def\baas{\reff@jnl{BAAS}}              
\def\jcap{\reff@jnl{JCAP}}          
\def\jrasc{\reff@jnl{JRASC}}            
\def\memras{\reff@jnl{MmRAS}}           
\def\mnras{\reff@jnl{MNRAS}}            
\def\newastro{\reff@jnl{New Astron.}}   
\def\pra{\reff@jnl{Phys.Rev.A}}         
\def\prb{\reff@jnl{Phys.Rev.B}}         
\def\prc{\reff@jnl{Phys.Rev.C}}         
\def\prd{\reff@jnl{Phys.Rev.D}}         
\def\prl{\reff@jnl{Phys.Rev.Lett}}      
\def\pasp{\reff@jnl{PASP}}              
\def\pasj{\reff@jnl{PASJ}}              
\def\qjras{\reff@jnl{QJRAS}}            
\def\skytel{\reff@jnl{S\&T}}            
\def\solphys{\reff@jnl{Solar~Phys.}}    
\def\sovast{\reff@jnl{Soviet~Ast.}}     
\def\ssr{\reff@jnl{Space~Sci.Rev.}}     
\def\zap{\reff@jnl{ZAp}}                
\def\nat{\reff@jnl{Nature}}             
\def\physrep{\reff@jnl{Phys.~Rep.}}     
\def\prog{\reff@jnl{PThPS}}        

\begin{document}

\title{Nonlinear stochastic biasing of halos:\\ Analysis of cosmological
  $N$-body simulations and perturbation theories}

\author{Masanori Sato$^{1}$\footnote{masanori@nagoya-u.jp} and
Takahiko Matsubara$^{1,2}$ 
}
\affiliation{%
$^{1}$ Department of Physics, Nagoya University, Chikusa, Nagoya 464--8602, Japan
}%
\affiliation{%
$^{2}$ Kobayashi-Maskawa Institute for the Origin of
 Particles and the Universe, Nagoya University, Chikusa, Nagoya 464--8602, Japan
}%

\date{\today}


\begin{abstract}
    It is crucial to understand and model a behavior of galaxy biasing
    for future ambitious galaxy redshift surveys. Using 40 large
    cosmological $N$-body simulations for a standard $\Lambda$CDM
    cosmology, we study the cross-correlation coefficient between
    matter and the halo density field, which is an indicator of
    the stochasticity of bias, over a wide redshift range $0\le z \le 3$.
    The cross-correlation coefficient is important to extract
    information on the matter density field, e.g., by combining galaxy
    clustering and galaxy-galaxy lensing measurements. We compare the
    simulation results with integrated perturbation theory (iPT)
    proposed by one of the present authors and standard perturbation
    theory (SPT) combined with a phenomenological model of local bias.
    The cross-correlation coefficient derived from the iPT agrees with
    $N$-body simulation results down to $r\sim$ 15 (10) $h^{-1}$Mpc
    within 0.5 (1.0) $\%$ for all redshifts and halo masses we
    consider. The SPT with local bias does not explain complicated
    behaviors on quasilinear scales at low redshifts, while roughly
    reproduces the general behavior of the cross-correlation coefficient
    on fully nonlinear scales. The iPT is powerful to predict the
    cross-correlation coefficient down to quasilinear regimes with a
    high precision.
\end{abstract}

\maketitle

\section{Introduction}

In the standard cosmological model, known as the $\Lambda$CDM model,
the energy density is dominated by mysterious components called
dark matter and dark energy. The correlation function of dark matter
and its Fourier counterpart, the power spectrum, contain a wealth of
information that can be used to determine, e.g., the dark matter, dark
energy, and neutrino masses. Thus, it is very important to exploit
these quantities in the large-scale structure of the universe, which
is a pillar of modern observational cosmology. However, how to take account of galaxy biasing needs to be investigated. Observable
galaxies are biased relative to the underlying matter density field.
The galaxy biasing is affected by nonlinear effects and is scale
dependent in general. Such nonlinear effects impose a serious problem
in analyzing galaxy surveys \citep[e.g.,][]{2006ApJ...645..977B,
  2007ApJ...657..645P, 2008MNRAS.385..830S, 2011MNRAS.415.2876B}.
Upcoming galaxy surveys such as
BigBOSS\footnote{http://bigboss.lbl.gov/}~\citep{2011arXiv1106.1706S},
Euclid~\citep{2011arXiv1110.3193L}, Subaru Prime Focus Spectrograph
(PFS)\footnote{http://sumire.ipmu.jp/pfs/intro.html}~\citep{2012arXiv1206.0737E},
and the Wide-Field Infrared Survey Telescope
(WFIRST)\footnote{http://wfirst.gsfc.nasa.gov/} require the
understanding of galaxy biasing with high precision and thus
a theoretically precise description of the galaxy biasing is a crucial
issue.

Most of the direct studies of the clustering of matter on cosmological
scales rely on shear-shear weak lensing, but it is also possible to
extract information on the matter clustering by combining galaxy
clustering and galaxy-galaxy lensing
measurements~\citep[e.g.,][]{2012arXiv1207.1120M}. To achieve this,
one has to precisely know the relation between the distribution of
galaxies and the distribution of matter. An important property of the
relation is often characterized by a cross-correlation coefficient.
The cross-correlation coefficient is a characteristic parameter of
stochasticity~\citep{1999ApJ...520...24D}. Since
galaxies are expected to form in dark matter halos in modern models of
galaxy formation, understanding and modeling the clustering properties
of the halos play an important role and are crucial first steps in
modeling galaxy biasing.

In this work, we examine how well-known models of halo clustering
reproduce the cross-correlation coefficient between matter and halo
density fields obtained from $N$-body simulations. We consider two
models of nonlinear bias: the integrated Perturbation Theory (iPT)
developed by \citet{2011PhRvD..83h3518M} which naturally incorporates
the halo bias, redshift-space distortions, nonlocal Lagrangian bias,
and primordial non-Gaussianity in a formalism of perturbation theory
with a resummation technique based on the Lagrangian picture (see also
\citep{2008PhRvD..77f3530M,2008PhRvD..78h3519M}), and the standard
perturbation theory (SPT) combined with the phenomenological model of
local bias, which leads to nontrivial renormalizations of the
leading-order bias parameter~\citep{2006PhRvD..74j3512M}. A significant
advantage of the iPT is that it is simpler and easier to use to calculate the
power spectrum than other resummation methods even in the presence of
halo bias and redshift-space distortions. The computational cost is
similar to that of the SPT.

We focus not on the power spectrum but on the two-point correlation function,
because we do not suffer from shot noise effect in the correlation
function. While two-loop corrections in the iPT generally might have an
impact on the correlation function on quasilinear
scales~\citep{2011JCAP...08..012O}, we use one-loop iPT for simplicity
in this paper.

This paper is organized as follows. We first review the theoretical
predictions of the power spectrum and correlation function in
Section~\ref{sec:ana}. We describe the details of $N$-body simulations
and a method to compute the correlation functions of matter and halos
from $N$-body simulations in Section~\ref{sec:Nbody}. After showing
the results of the matter and halo correlation functions and its
cross-correlation function in Section~\ref{sec:corr}, we then show the
main results of this paper in Section~\ref{sec:stocha}. Finally,
Section~\ref{sec:conc} is devoted to our conclusion.

\section{Analytic Models}
\label{sec:ana}

In this section, we briefly review two theoretical models: the iPT model
with nonlocal bias and the SPT model with local bias, which are
compared with $N$-body simulation results.

\subsection{Predictions of integrated Perturbation Theory}

We use iPT~\citep{2011PhRvD..83h3518M} to investigate how the
cross-correlation coefficient behaves on quasilinear scales for
various halo masses and redshifts. It is convenient to write down the
power spectrum predictions of the iPT based on multipoint propagators
recently introduced in \citet{2008PhRvD..78j3521B}. Using the
multipoint propagators $\varGamma^{(n)}$, the one-loop power spectrum
between object $X$ and $Y$ based on the iPT can be written as (the
full derivation is given in \citep{Matsubara2013})
\begin{widetext}
\begin{equation}
  P_{XY}(k) = \varPi^{2}(k)
  \left[
    \hat{\varGamma}_X^{(1)}(k)\varGamma_Y^{(1)}(k) P_{\rm L}(k)
    + \frac{k^3}{8\pi^2}\int_0^{\infty}\pd r\,r^2\int_{-1}^{1}\pd x\,
    \hat{\varGamma}_X^{(2)}(k,r,x) \hat{\varGamma}_Y^{(2)}(k,r,x) P_{\rm L}(kr)
    P_{\rm L}\left(k\sqrt{1+r^2-2rx}\right)
  \right],
\end{equation}
where indices $X$ and $Y$ are either matter 'm' or halo 'h' in this
paper, $P_{\rm L}(k)$ is the linear matter power spectrum, and the vertex
factor $\varPi(k)$ is given by
\begin{equation}
 \varPi(k)=\exp\left[-\frac{k^2}{12\pi^2}\int\pd p\,P_{\rm L}(p)\right].
\end{equation}
The normalized multipoint propagators $\hat{\varGamma}^{(1)}$ and
$\hat{\varGamma}^{(2)}$ are given by
\begin{align}
  \hat{\varGamma}_{\rm m}^{(1)}(k) &=
  1 + \frac{k^3}{4\pi}\int_0^{\infty}\pd r\,
  \int_{-1}^{1}\pd x\,
  \left[
    \frac{5}{21}\frac{r^2(1-x^2)^2}{1+r^2-2rx}
    + \frac{3}{7}\frac{(1-rx)(1-x^2)rx}{1+r^2-2rx}
  \right] P_{\rm L}(kr),
\label{Gamma_m1}\\
  \hat{\varGamma}_{\rm h}^{(1)}(k) &=
  1 + c^{(1)}_{\rm h}(k)
  + \frac{k^3}{4\pi} \int_0^{\infty}\pd r\,\int_{-1}^{1}\pd x\,
  \left\{
    \frac{5}{21}\frac{r^2(1-x^2)^2}{1+r^2-2rx}
    + \frac{3}{7}\frac{(1-rx)(1-x^2)}{1+r^2-2rx}
    \left[rx+r^2c^{(1)}_{\rm h}(kr)\right]
  \right\}
  P_{\rm L}(kr),
\label{Gamma_h1}\\
  \hat{\varGamma}_{\rm m}^{(2)}(k,r,x) &=
  \frac{x}{r} + \frac{1-rx}{1+r^2-2rx}
  - \frac{4}{7}\frac{1-x^2}{1+r^2-2rx},
\label{Gamma_m2}\\
  \hat{\varGamma}_{\rm
    h}^{(2)}(k,r,x)&=\frac{x}{r}
  \left[
    1+c^{(1)}_{\rm h}\left(k\sqrt{1+r^2-2rx}\right)
  \right]
  + \frac{1-rx}{1+r^2-2rx}
  \left[
    1+c^{(1)}_{\rm h}(kr)
  \right] - \frac{4}{7}\frac{1-x^2}{1+r^2-2rx} +
  c^{(2)}_{\rm h}\left(kr,k\sqrt{1+r^2-2rx}\right)
\label{Gamma_h2},
\end{align}
\end{widetext}
where indices 'm' and 'h' denote the matter and halo, and we assume that
the second-order renormalized bias function $c^{(2)}_{\rm
  h}(\bm{k}_1,\bm{k}_2)$ depends only on the magnitudes of the 
wave vectors, $k_1 \equiv |\bm{k}_1|$ and $k_2 \equiv |\bm{k}_2|$. From
Equations~(\ref{Gamma_m1})-(\ref{Gamma_h2}), we can easily understand
that the matter result is recovered when $c^{(1)}_{\rm h} =
c^{(2)}_{\rm h} = 0$. Here $c^{(1)}_{\rm h}$ and $c^{(2)}_{\rm h}$ are
renormalized bias functions in Lagrangian space introduced by
\citet{2011PhRvD..83h3518M} and obtained as
\begin{align}
  c^{(n)}_{\rm h}(\bm{k}_1,\dots,\bm{k}_n) =
  \frac{\displaystyle
    \int_{\nu_1}^{\nu_2}\frac{f_{\rm MF}(\nu)}{M}
    \hat{c}^{(n)}_{\rm h}(\bm{k}_1,\dots,\bm{k}_n;\nu)\pd\ln\nu}
       {\displaystyle
         \int_{\nu_1}^{\nu_2}\frac{f_{\rm MF}(\nu)}{M}\pd\ln\nu},
\end{align}
for a mass range $M_1\le M\le M_2$ (see Equations 64 and 108 of
\citet{2012PhRvD..86f3518M}), where $\nu=\delta_c/\sigma(M)$ is a
function of mass $M$, and $\delta_c$ is the critical overdensity for
spherical collapse. In an Einstein-de Sitter cosmology, the critical
overdensity is $\delta_c\approx{1.686}$, while it shows weak
dependence on cosmology and redshifts in general
cosmology~\citep{1997PThPh..97...49N,2000ApJ...534..565H}, and thus we
use the fitting formula introduced by \citet{2000ApJ...534..565H} to include
cosmological dependence. The function $\sigma(M)$ is the
root-mean-square linear density field smoothed with a top-hat filter
of radius $R$ enclosing an average mass $M=\rho_0 4\pi R^3/3$,
\begin{equation}
\sigma^2(M)=\int\frac{k^2\pd{k}}{2\pi^2}W^2(kR)P_{\rm L}(k),
\end{equation}
with
\begin{equation}
W(x)=\frac{3j_1(x)}{x}=\frac{3}{x^3}(\sin x-x\cos x),
\end{equation}
where $\rho_0$ is the mean matter density of the universe and $j_1(x)$
is the first-order spherical Bessel function.
$f_{\rm MF}(\nu)$ is the scaled differential mass function defined as~\citep{2001MNRAS.321..372J}
\begin{equation}
 f_{\rm MF}(\nu)=\frac{M}{\rho_0}n(M)\frac{\pd M}{\pd\ln\nu},
\end{equation}
where $n(M)$ is the comoving number density of halos with mass $M$.
The quantity $f_{\rm MF}(\nu)$ is frequently used in the literature
and there have been several analytic predictions
\citep{1974ApJ...187..425P,1991ApJ...379..440B,2001MNRAS.323....1S}
and fitting formulas \citep[e.g.,][]{1999MNRAS.308..119S,
  2001MNRAS.321..372J, 2006ApJ...646..881W, 2007MNRAS.374....2R,
  2010MNRAS.403.1353C, 2010MNRAS.402..589M, 2011ApJ...732..122B}.
In this paper, we use the fitting formula for the mass function introduced
 by \citet{2011ApJ...732..122B}, which shows better agreement with our
simulations~\citep{2011PhRvD..84d3501S}. $\hat{c}_n^{\rm L}$ is given
as~(see, Equations 92, 95, and 96 in \citet{2012PhRvD..86f3518M})
\begin{multline}
  \hat{c}^{(n)}_{\rm h}(\bm{k}_1,\dots,\bm{k}_n;\nu)
  = b_n^{\rm L}(M)W(k_1R)\cdots W(k_nR)
\\
  + \frac{A_{n-1}(M)}{\delta_c^n}
  \frac{\pd}{\pd\ln\sigma(M)}\left[W(k_1R)\cdots W(k_nR)\right],
\end{multline}
with
\begin{align}
  A_0(M) &= 1,
\\
  A_1(M) &= 1 + \delta_c b_1^{\rm L}(M),
\end{align}
where $b_n^{\rm L}$ is the Lagrangian bias function for the halo bias.

The theoretical two-point correlation function can be expressed in terms of
the power spectrum as
\begin{equation}
  \xi_{XY}(r) =
  \int \frac{k^2\pd k}{2\pi^2} \frac{\sin{(kr)}}{kr} P_{XY}(k).
\end{equation}

\subsection{Standard perturbation theory with local bias model}

In the SPT formalism, we consider the {\it local deterministic
  nonlinear biasing} model. Following \citet{1993ApJ...413..447F}, we
restrict the consideration on large scales in Eulerian space and
assume that the halo density can be described by a smoothed function
$\mathcal{F}(\delta_{\rm m})$ that depends only on the matter density.
We can expand $\mathcal{F}$ in a Taylor series around $\delta_{\rm m}$
such that
\begin{equation}
 \delta_{\rm h}=\mathcal{F}(\delta_{\rm
  m})=\sum_{n=1}^{\infty}\frac{b_n^{\rm E}}{n!}\delta_{\rm m}^{n},
\end{equation}
where $\delta_{\rm m}$ is the nonlinear matter density. We then
combine this expansion with SPT, which expands the matter density
perturbations into a series $\delta_{\rm m}=\delta_{\rm
  m}^{(1)}+\delta_{\rm m}^{(2)}+\cdots$, where $\delta_{\rm m}^{(1)}$
is the linear density field and $\delta_{\rm m}^{(n)}$ is of order
$[\delta_{\rm m}^{(1)}]^n$. At the next-to-leading order, we can
obtain the auto- and cross-power spectrum of halos
as~\citep{2006PhRvD..74j3512M,2010PhRvD..81f3531B}
\begin{align}
 P_{\rm hh}(k)&=b_1^2 P_{\rm NL}(k)+2b_1b_2A(k)+\frac{b_2^2}{2}B(k)+N,\\
 P_{\rm hm}(k)&=b_1 P_{\rm NL}(k)+b_2A(k),
\end{align}
where $b_1$ and $b_2$ are the renormalized bias parameters, $N$ is the
renormalized shot noise, and $P_{\rm NL}(k)$ is the nonlinear matter
power spectrum. $b_1$ and $b_2$ should be determined empirically or
treated as free parameters. In this paper, we will examine both cases
in Section~\ref{sec:stocha}. The terms $A(k)$ and $B(k)$ can be
obtained as
\begin{align}
 A(k)&=\int\frac{\pd^3 q}{(2\pi)^3}F_2(\bm{q},\bm{k}-\bm{q})P_{\rm
 L}(q)P_{\rm L}(|\bm{k}-\bm{q}|),\\
 B(k)&=\int\frac{\pd^3 q}{(2\pi)^3}P_{\rm L}(q)\left[P_{\rm
 L}(|\bm{k}-\bm{q}|)-P_{\rm L}(q)\right],
\end{align}
where $F_2$ is the second-order mode coupling kernel in SPT,
\begin{equation}
  F_2(\bm{k}_1,\bm{k}_2) =
  \frac{5}{7} + \frac{1}{2}\frac{\bm{k}_1\cdot\bm{k}_2}{k_1k_2}
  \left(\frac{k_1}{k_2} + \frac{k_2}{k_1}\right)
  + \frac{2}{7} \left(\frac{\bm{k}_1\cdot\bm{k}_2}{k_1k_2}\right)^2.
\end{equation}

Taking Fourier transforms, we then obtain corresponding correlation
functions given by
\begin{align}
 \xi_{\rm hh}(r) &=
 b_1^2 \xi_{\rm NL}(r) + 2b_1b_2A(r) + \frac{b_2^2}{2}B(r),
\label{spthh}\\
 \xi_{\rm hm}(r) &= b_1 \xi_{\rm NL}(r) + b_2A(r),
\label{spthm}
\end{align}
where $\xi_{\rm NL}$ is the nonlinear matter correlation function, and
$A(r)$ and $B(r)$ are the Fourier transforms of $A(k)$ and $B(k)$. Note
that $B(r)=\xi_{\rm L}^2(r)-\sigma_{\rm c}^2\,\delta_{D}(\bm{r})$
where $\xi_{\rm L}(r)$ is the linear matter correlation function,
$\sigma_{\rm c}^2=\int\pd^3 q\,P^2_{\rm L}(q)/(2\pi)^3$, and
$\delta_{D}(\bm{r})$ is the Dirac delta function.

\section{$N$-body simulations}
\label{sec:Nbody}
\subsection{Simulation parameters}

\begin{table*}
\caption{
Parameters in high- and low-resolution $N$-body simulations: the matter
 density $\Omega_{\rm m}$, the dark energy density $\Omega_{\Lambda}$,
 the baryon density $\Omega_{\rm b}$, the Hubble parameter $h$, the
 spectral index $n_s$, the variance of the density perturbations at
 8$h^{-1}$Mpc $\sigma_8$, the box size $L_{\rm box}$, the number of
 particles $N_p$, the initial redshift $z_{\rm ini}$, the softening
 length $r_s$, and the number of realizations $N_{\rm run}$.
}
\label{table1}
\begin{center}
\begin{tabular}{p{35mm}ccccccccccc}
\hline\hline  
Name & $\Omega_{\rm m}$ & $\Omega_{\Lambda}$ &
 $\Omega_{\rm b}$ & $h$ & $n_s$ & $\sigma_8$ & $L_{\rm box}$ & $N_p$ &
 $z_{\rm ini}$ & $r_{\rm s}$ & $N_{\rm run}$\\ \hline
L1000 (high resolution) & 0.265 & 0.735 & 0.0448 & 0.71 & 0.963 & 0.80 & 1000$h^{-1}$Mpc &
				 1024$^3$ & 36 & 50$h^{-1}$kpc & 30\\
L2000 (low resolution) & 0.265 & 0.735 & 0.0448 & 0.71 & 0.963 & 0.80 & 2000$h^{-1}$Mpc &
				 1024$^3$ & 31 & 100$h^{-1}$kpc & 10\\ \hline\hline
\end{tabular}
\end{center}
\end{table*}

\begin{table*}[htb]
\caption{ Properties of halo catalogs of high- and low-resolution
  $N$-body simulations for each mass bin. We use the five halo catalogs
  abbreviated as ``Bin 1'',$\dots$, ``Bin 5''. $\bar{N}_{\rm h}$ and
  $\bar{M}_{\rm h}$ are the average halo numbers and average halo
  masses at various redshifts.
}
\label{table2}
\begin{center}
\begin{tabular}[c]{ccc|ccc|ccc}
\multicolumn{9}{c}{\bf L1000}\\ \hline\hline
\multicolumn{3}{c|}{Bin 1}&
\multicolumn{3}{c|}{Bin 2}&
\multicolumn{3}{c}{Bin 3}\\
\multicolumn{3}{c|}{$1.37\le M_{\rm h}/(10^{12}{h}^{-1}M_{\odot})<4.11$}&
\multicolumn{3}{c|}{$4.11\le M_{\rm h}/(10^{12}{h}^{-1}M_{\odot})<12.32$}&
\multicolumn{3}{c}{$1.23\le M_{\rm h}/(10^{13}{h}^{-1}M_{\odot})<3.70$}\\
\hline  
 $\quad z\quad$ & $\bar{N}_{\rm h}$ & $\bar{M}_{\rm h}[h^{-1}M_{\odot}]$ &
 $\quad z\quad$ & $\bar{N}_{\rm h}$ & $\bar{M}_{\rm h}[h^{-1}M_{\odot}]$ & 
 $\quad z\quad$ & $\bar{N}_{\rm h}$ & $\bar{M}_{\rm h}[h^{-1}M_{\odot}]$
\\ \hline
 $\quad 3.0\quad$ & 3.56$\times{10}^5$ &
 2.08$\times{10}^{12}$ & $\quad 3.0\quad$ & 4.14$\times{10}^4$ &
 6.06$\times{10}^{12}$ & $\quad 3.0\quad$ & 2.50$\times{10}^3$ &
 1.69$\times{10}^{13}$\\ 
 $\quad 2.0\quad$ & 9.88$\times{10}^5$ &
 2.17$\times{10}^{12}$ & $\quad 2.0\quad$ & 1.97$\times{10}^5$ &
 6.42$\times{10}^{12}$ & $\quad 2.0\quad$ & 2.75$\times{10}^4$ &
 1.83$\times{10}^{13}$\\
 $\quad 1.0\quad$ & 1.73$\times{10}^6$ &
 2.23$\times{10}^{12}$ & $\quad 1.0\quad$ & 4.98$\times{10}^5$ &
 6.69$\times{10}^{12}$ & $\quad 1.0\quad$ & 1.24$\times{10}^5$ &
 1.96$\times{10}^{13}$\\
 $\quad 0.5\quad$ & 1.95$\times{10}^6$ &
 2.24$\times{10}^{12}$ & $\quad 0.5\quad$ & 6.25$\times{10}^5$ &
 6.78$\times{10}^{12}$ & $\quad 0.5\quad $ & 1.89$\times{10}^5$ &
 2.00$\times{10}^{13}$\\
 $\quad 0.3\quad$ & 1.99$\times{10}^6$ &
 2.25$\times{10}^{12}$ & $\quad 0.3\quad$ & 6.60$\times{10}^5$ &
 6.81$\times{10}^{12}$ & $\quad 0.3\quad$ & 2.12$\times{10}^5$ &
 2.01$\times{10}^{13}$\\
 $\quad 0\quad$ & 2.02$\times{10}^6$ &
 2.25$\times{10}^{12}$ & $\quad 0\quad$ & 6.94$\times{10}^5$ &
 6.84$\times{10}^{12}$ & $\quad 0\quad$ & 2.39$\times{10}^5$ &
 2.03$\times{10}^{13}$\\\hline\hline\\
\end{tabular}
\begin{tabular}[c]{ccc|ccc}
\multicolumn{6}{c}{\bf L2000}\\ \hline\hline
\multicolumn{3}{c|}{Bin 4}&
\multicolumn{3}{c}{Bin 5}\\
\multicolumn{3}{c|}{$1.23\le M_{\rm h}/(10^{13}{h}^{-1}M_{\odot})<3.70$}&
\multicolumn{3}{c}{$3.70\le M_{\rm h}/(10^{13}{h}^{-1}M_{\odot})<11.09$}\\
\hline  
 $\quad z\quad$ & $\bar{N}_{\rm h}$ & $\bar{M}_{\rm h}[h^{-1}M_{\odot}]$ &
 $\quad z\quad$ & $\bar{N}_{\rm h}$ & $\bar{M}_{\rm h}[h^{-1}M_{\odot}]$
\\ \hline
 $\quad 3.0\quad$ & 2.09$\times{10}^4$ &
 1.70$\times{10}^{13}$ & $\quad 3.0\quad$ & 3.85$\times{10}^3$ &
 4.70$\times{10}^{13}$\\ 
 $\quad 2.0\quad$ & 2.30$\times{10}^5$ &
 1.84$\times{10}^{13}$ & $\quad 2.0\quad$ & 1.60$\times{10}^4$ &
 5.17$\times{10}^{13}$\\
 $\quad 1.0\quad$ & 1.05$\times{10}^6$ &
 1.97$\times{10}^{13}$ & $\quad 1.0\quad$ & 1.88$\times{10}^5$ &
 5.70$\times{10}^{13}$\\
 $\quad 0.5\quad$ & 1.61$\times{10}^6$ &
 2.02$\times{10}^{13}$ & $\quad 0.5\quad$ & 3.96$\times{10}^5$ &
 5.92$\times{10}^{13}$\\
 $\quad 0.3\quad$ & 1.80$\times{10}^6$ &
 2.03$\times{10}^{13}$ & $\quad 0.3\quad$ & 4.89$\times{10}^5$ &
 5.99$\times{10}^{13}$\\
 $\quad 0\quad$ & 2.03$\times{10}^6$ &
 2.04$\times{10}^{13}$ & $\quad 0\quad$ & 6.17$\times{10}^5$ &
 6.07$\times{10}^{13}$\\\hline\hline
\end{tabular}
\end{center}
\end{table*}

To obtain accurate predictions of the cross-correlation coefficient,
we resort to the use of high-resolution $N$-body simulations of structure
formation. To perform the $N$-body simulations, we use a publicly
available tree-particle mesh code, {\em
  Gadget2}~\citep{2005MNRAS.364.1105S}. We adopt the standard
$\Lambda$CDM model with the matter density $\Omega_{\rm m}=0.265$, the
baryon density $\Omega_{\rm b}=0.0448$, the dark energy density
$\Omega_{\Lambda}=0.735$ with equation of state parameter $w=-1$, the
spectral index $n_s=0.963$, the variance of the density perturbations
in a sphere of radius 8$h^{-1}$Mpc $\sigma_8=0.80$, and the Hubble
parameter $h=0.71$. These cosmological parameters are consistent with
the Wilkinson Microwave Anisotropy Probe 7-year
results~\citep{2011ApJS..192...18K}. We performed two types of
simulations, both with $N_p=1024^3$ particles in cubic boxes. The first
type has a side $L_{\rm box}=1000h^{-1}$Mpc with softening length
$r_{\rm s}$ being $50h^{-1}$kpc, and the second type has a side
$L_{\rm box}=2000h^{-1}$Mpc with softening length $r_{\rm s}$ being
$100h^{-1}$kpc. These two types are named as L1000 and L2000,
respectively. The initial conditions are generated based on the
second-order Lagrangian perturbation
theory~\citep{2006MNRAS.373..369C,2011A&A...527A..87V} with the
initial linear power spectrum calculated by {\sc
  CAMB}~\citep{2000ApJ...538..473L}. The initial redshift is set to
$z_{\rm ini}=36$ for L1000 and $z_{\rm ini}=31$ for L2000. We perform
$N_{\rm run}=30$ and 10 realizations for L1000 and L2000,
respectively. We summarize the simulation parameters in
Table~\ref{table1}. The L1000 simulations used in this paper are the
same as L1000 used in \citet{2011PhRvD..84d3501S}.

We store outputs at $z=3.0$, 2.0, 1.0, 0.5, 0.3, and 0 and identify halos for
each output using a Friends-of-Friends (FOF) group finder with linking length
of 0.2 times the mean separation~\citep{1985ApJ...292..371D}. 
We select halos in which the number of particles, $N_p$, is equal to or
larger than 20 which corresponds to the halos with masses $1.37\times
10^{12}h^{-1}M_{\odot}$ for L1000 and $1.10\times
10^{13}h^{-1}M_{\odot}$ for L2000.
Then we divide halos into five mass bins to keep track of their
different clustering properties.
The average number and mass of halos among realizations for
redshifts 
are listed in Table~\ref{table2}. The halo catalogs of Bin 4 in L2000
is
constructed so that the halo mass range is the same as that of Bin
3 in L1000, as shown in Table~\ref{table2}.
Since the volume of L2000 simulations is bigger than that of L1000 simulations,
the number of halos with a certain mass are larger for L2000 simulations.

\subsection{Analysis: two-point correlation functions}

To calculate the two-point correlation function of dark matter from
$N$-body simulations, we first randomly choose the number of particles
$N_{p,r}=196^3$ and $256^3$ for L1000 and L2000. For dark matter
halos, we use all halos in each bin. Then we directly count the
$N$-body particle and/or halos to calculate the two-point correlation
function instead of using the fast Fourier
transform method~\citep{2011PhRvD..84d3501S}. We choose $r_i$ to be the
center of the $i$th bin, i.e., $r_i=(r_i^{\rm min}+r_i^{\rm max})/2$,
where $r_i^{\rm min}$ and $r_i^{\rm max}$ are the minimum and maximum
distances of the $i$th bin.

The shot noise corrections in the halo power spectrum are subtle. If
the dark matter halos are regarded as a Poisson process, we can easily
subtract the shot noise effect by using the number density of halos $\bar{n}_h$.
However, \citet{2007PhRvD..75f3512S} found that this standard
correction method is not exactly correct for halos, particularly for
those of large mass. This is probably because in order to identify
halos using the FOF algorithm, we automatically impose that distances
between halos are larger than the sum of their radii, or they would
have been linked as bigger halos. Thus, the shot noise effect is scale
dependent and it is difficult to correctly subtract the effect of shot
noise. Therefore, we use the correlation function instead of using the
power spectrum, because the shot noise effect in the correlation
function is weaker than that in the power spectrum.


\section{Correlation functions}\label{sec:corr}

\begin{figure}
\begin{center}
 \includegraphics[width=0.45\textwidth]{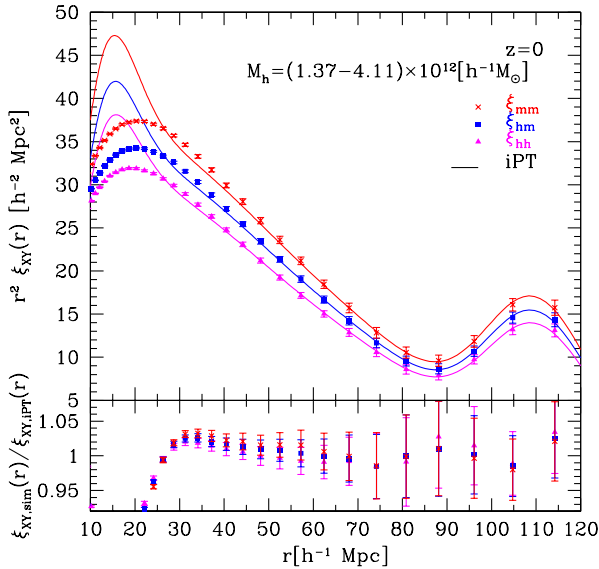}
\end{center}
\vskip-\lastskip
\caption{ {\it Top panel}: Correlation functions of matter, halo, and
  their cross-correlation function at redshift $z=0$, multiplied by a
  factor of $r^2$. For the halo mass range, we consider $1.37\le M_{\rm
    h}/(10^{12}{h}^{-1}M_{\odot})<4.11$ (Bin 1). The symbols are the
  results obtained from $N$-body simulations. The solid lines are the
  results of integrated perturbation theory
  (iPT)~\citep{2011PhRvD..83h3518M}. {\it Bottom panel}: Ratio of
  correlation functions measured from $N$-body simulations to those
  from the iPT. }
\label{fig:cor}
\end{figure}

Before presenting the results for the cross-correlation coefficient, we
compare the $N$-body simulation results with the iPT for the correlation
functions themselves.

Figure~\ref{fig:cor} shows the results for the correlation functions
of matter and halos, and their cross-correlation function at $z=0$. We
use the halo catalog of ``Bin 1'' shown in Table~\ref{table1}. The
amplitude of the halo-halo correlation is smaller than that of
the matter-matter correlation, because the halo bias $b_1^{\rm E}$ in this
halo range is 0.904 (less than 1). The error bars describe the
1-$\sigma$ error on the mean values obtained from 30 realizations. The
error bars increase on large scales because of the finite size of the
simulation box. The iPT predictions agree with $N$-body simulation
results down to $r\sim$ 25$h^{-1}$Mpc within a few percent for all
correlations. In Section~\ref{sec:stocha}, we will see that a range of
a few percent-level agreement in the cross-correlation coefficient is
extended more than that in the correlation functions.

\section{Cross-Correlation Coefficient}\label{sec:stocha}

\begin{figure*}
\begin{center}
 \includegraphics[width=0.9\textwidth]{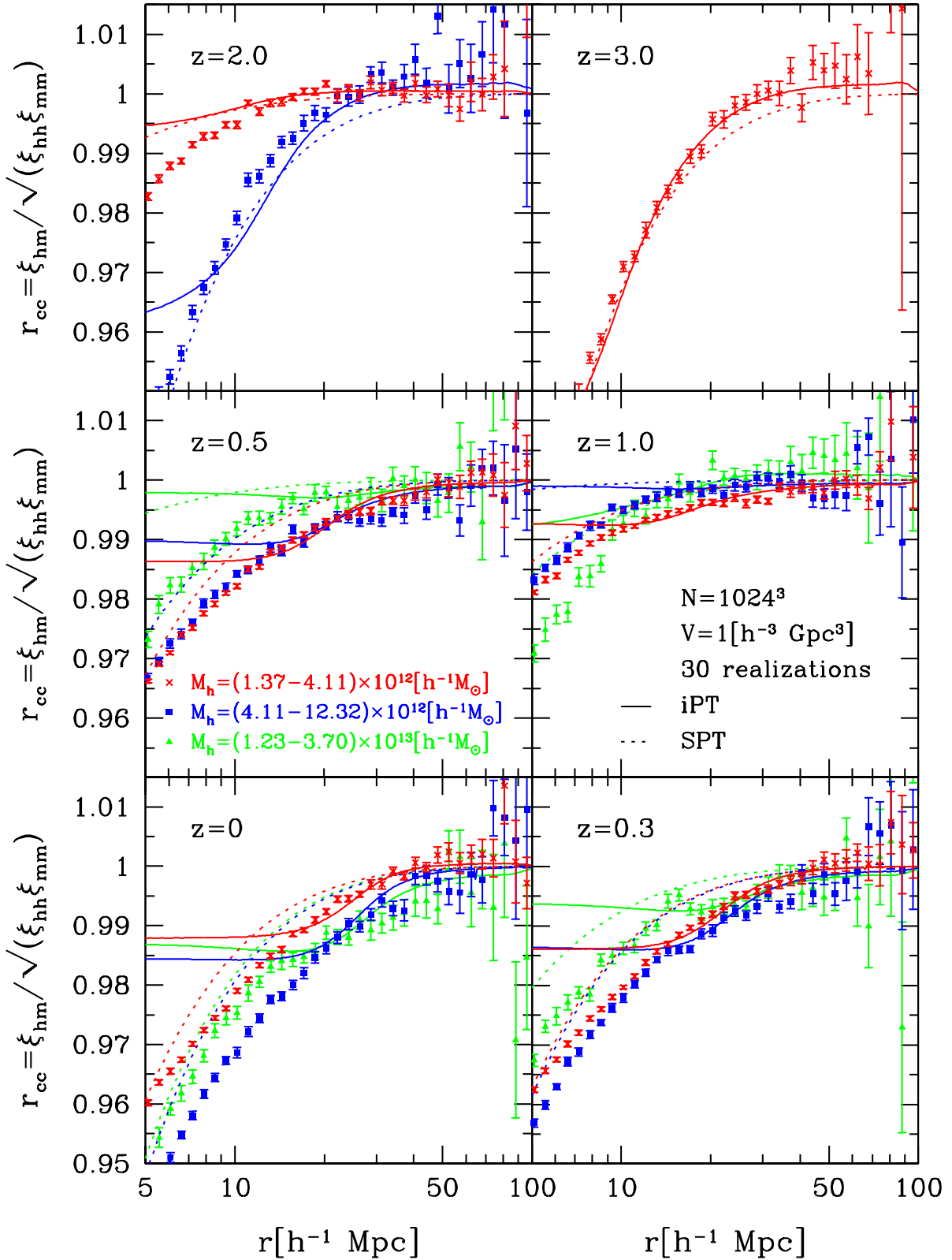}
\end{center}
\vskip-\lastskip
\caption{ The cross-correlation coefficient between the matter and halo
  density fields at $z=0$, 0.3, 0.5, 1.0, 2.0, and 3.0. For the halo mass
  ranges, we consider $1.37\le M_{\rm
    h}/(10^{12}{h}^{-1}M_{\odot})<4.11$ (Bin 1), $4.11\le M_{\rm
    h}/(10^{12}{h}^{-1}M_{\odot})<12.32$ (Bin 2), and $1.23\le M_{\rm
    h}/(10^{13}{h}^{-1}M_{\odot})<3.70$ (Bin 3). We do not plot the
  results in which the sum of the 1-$\sigma$ error bars in a range of
  $5\le r_i/(h^{-1}{\rm Mpc})\le 100$ is larger than 0.12, i.e.,
  $\sum_{5\le r_i\le 100}\sigma_i>0.12$.
The symbols are the results measured from $N$-body simulations. The
solid lines are the results of integrated perturbation theory
(iPT)~\citep{2011PhRvD..83h3518M} while the dotted lines are the
results of the simple model derived from standard perturbation theory with
local bias model (Equation~\ref{cc_bal}).
To empirically estimate $b_1$ and $b_2$, we use the relation in
 Equations~(\ref{bias_sph1}) and (\ref{bias_sph2}) and then simply substitute
 $b_1^{\rm E}$ and $b_2^{\rm E}$ with $b_1$ and $b_2$.
}
\label{fig:cc.zv3}
\end{figure*}

\begin{figure*}
\begin{center}
 \includegraphics[width=0.9\textwidth]{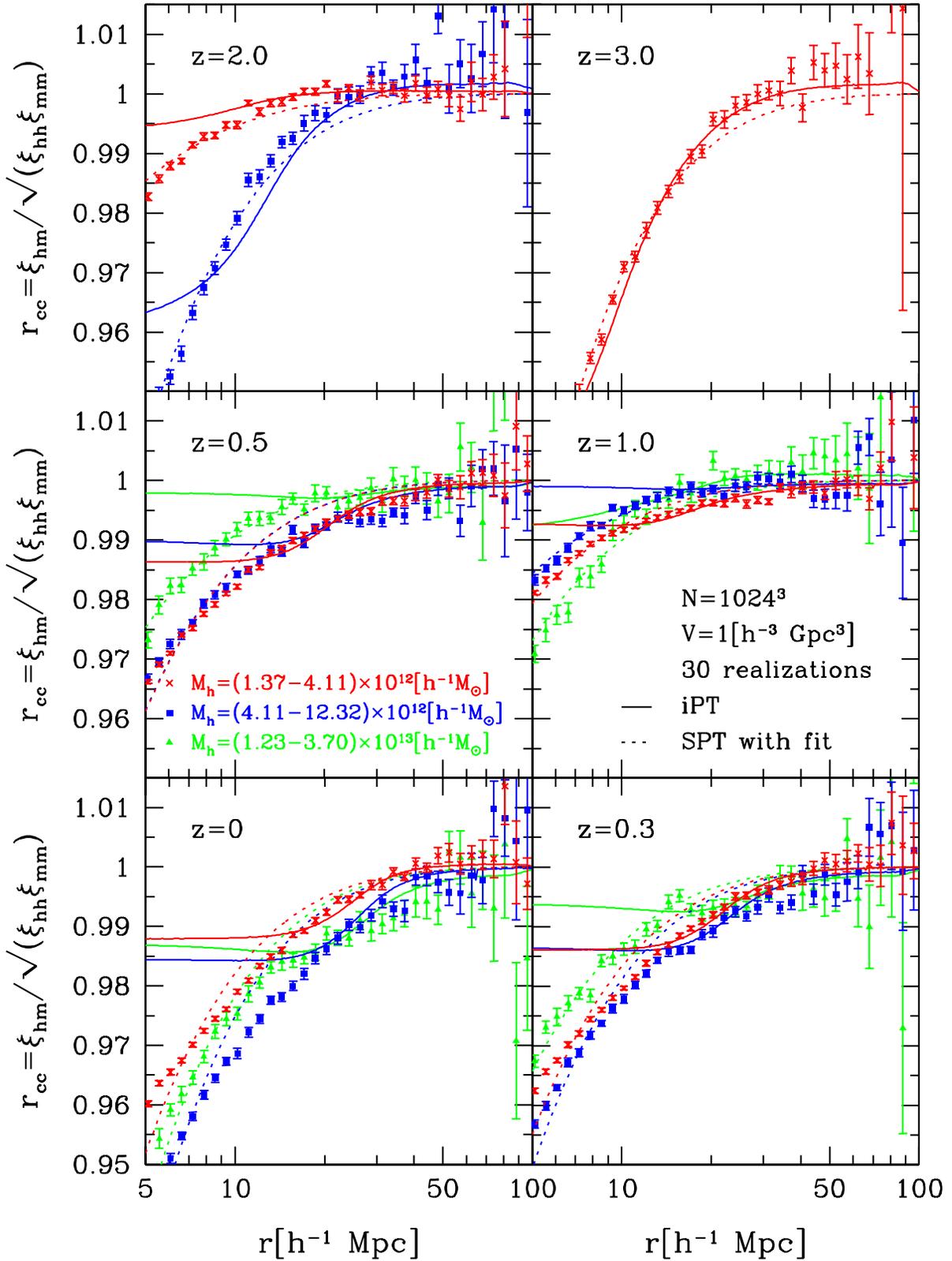}
\end{center}
\vskip-\lastskip
\caption{
Same as Figure~\ref{fig:cc.zv3}, but for the dotted lines, we fit $b_2/b_1$ to the $N$-body
 simulation results using a chi-square fit.
}
\label{fig:cc.zfit}
\end{figure*}

\begin{figure*}
\begin{center}
 \includegraphics[width=0.9\textwidth]{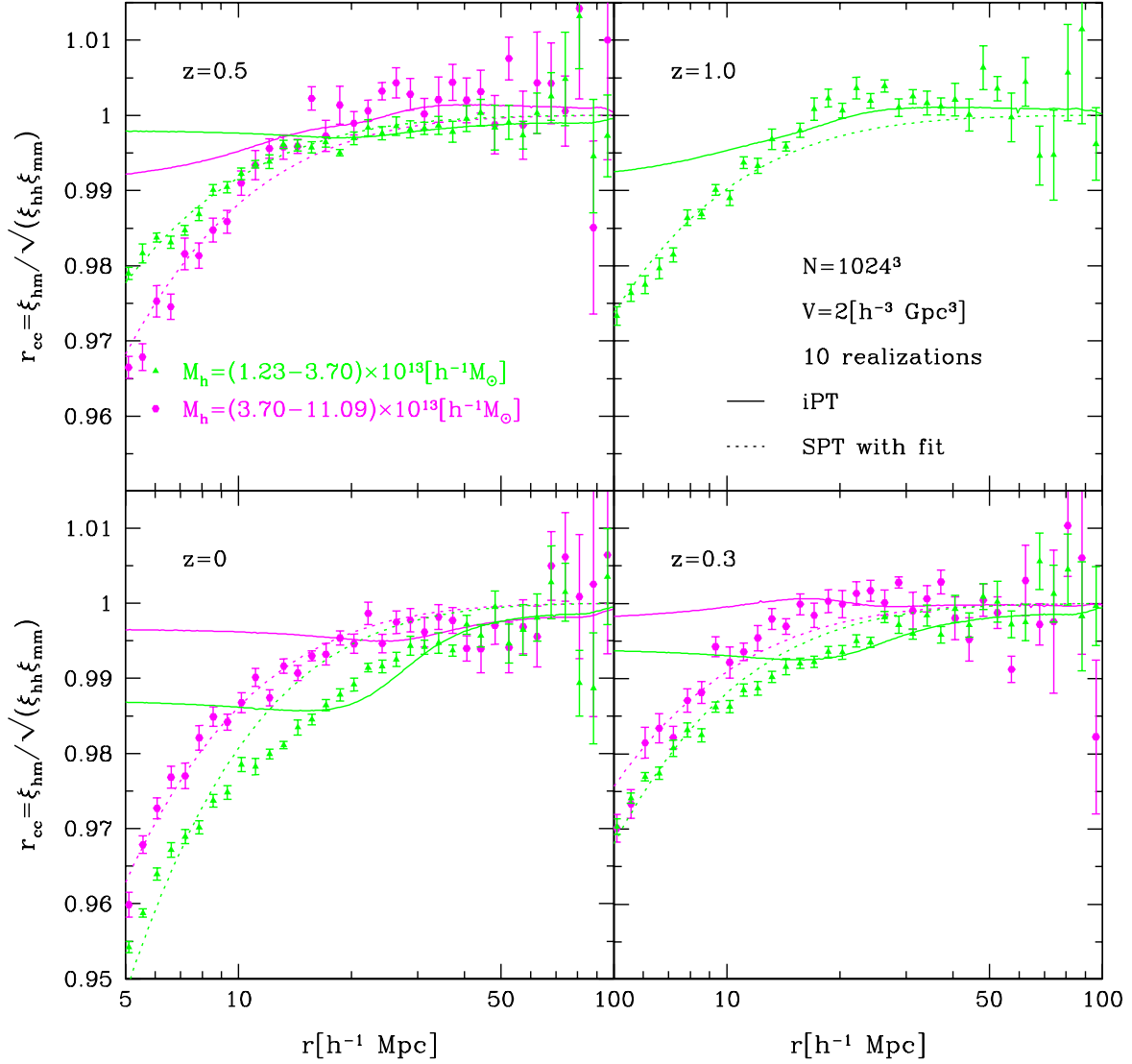}
\end{center}
\vskip-\lastskip
\caption{ The cross-correlation coefficients with halo mass ranges
  $1.23\le M_{\rm h}/(10^{13}{h}^{-1}M_{\odot})<3.70$ (Bin 4) and
  $3.70\le M_{\rm h}/(10^{13}{h}^{-1}M_{\odot})<11.09$ (Bin 5), given
  at redshift $z=0$, 0.3, 0.5, and 1.0. We do not show the results in
  which the sum of the 1-$\sigma$ error bars in a range of $5\le
  r_i/(h^{-1}{\rm Mpc})\le 100$ is larger than 0.12. The triangle and
  circle symbols are the results of $N$-body simulations. The solid
  lines correspond to the results of integrated perturbation theory
  (iPT)~\citep{2011PhRvD..83h3518M} while the dotted lines correspond
  to results of standard perturbation theory with the fitted bias model
  (Equation~\ref{cc_bal}).
}
\label{fig:cc.zfith}
\end{figure*}

\begin{figure*}
\begin{center}
 \includegraphics[width=0.95\textwidth]{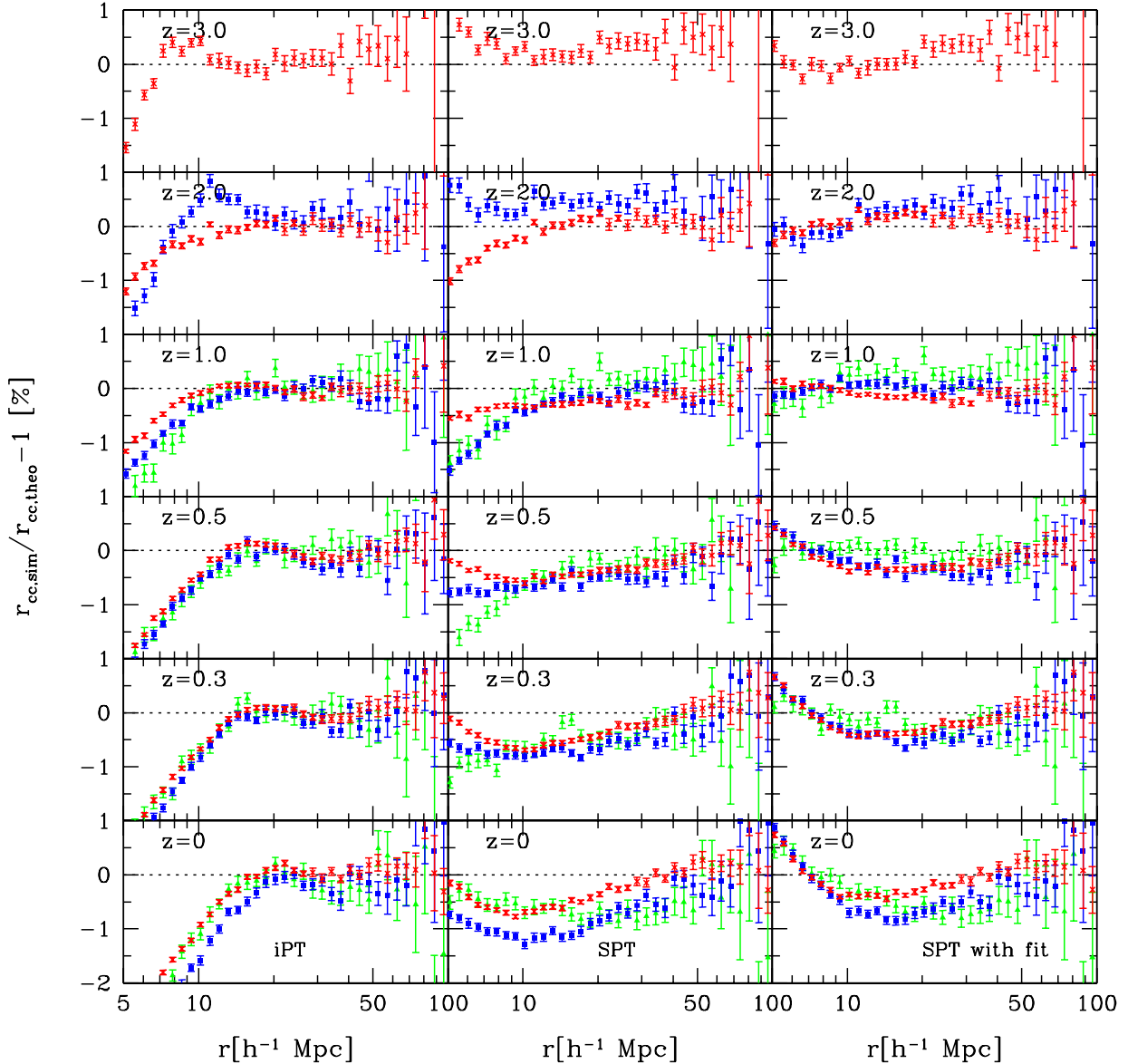}
\end{center}
\vskip-\lastskip
\caption{ Fractional differences between $N$-body results and
  theoretical predictions are shown in percents, at
  redshifts $z=0$, 0.3, 0.5, 1.0, 2.0, and 3.0, from {\it bottom} to
  {\it top}. {\it Left panels:} iPT, {\it Middle panels:} SPT, and
  {\it Right panels:} SPT with fitting.
 The red cross, blue box, and green triangles are the results of Bin 1, 2, and 3.
}
\label{fig:cc.ratio}
\end{figure*}

\begin{figure*}
\begin{center}
 \includegraphics[width=0.95\textwidth]{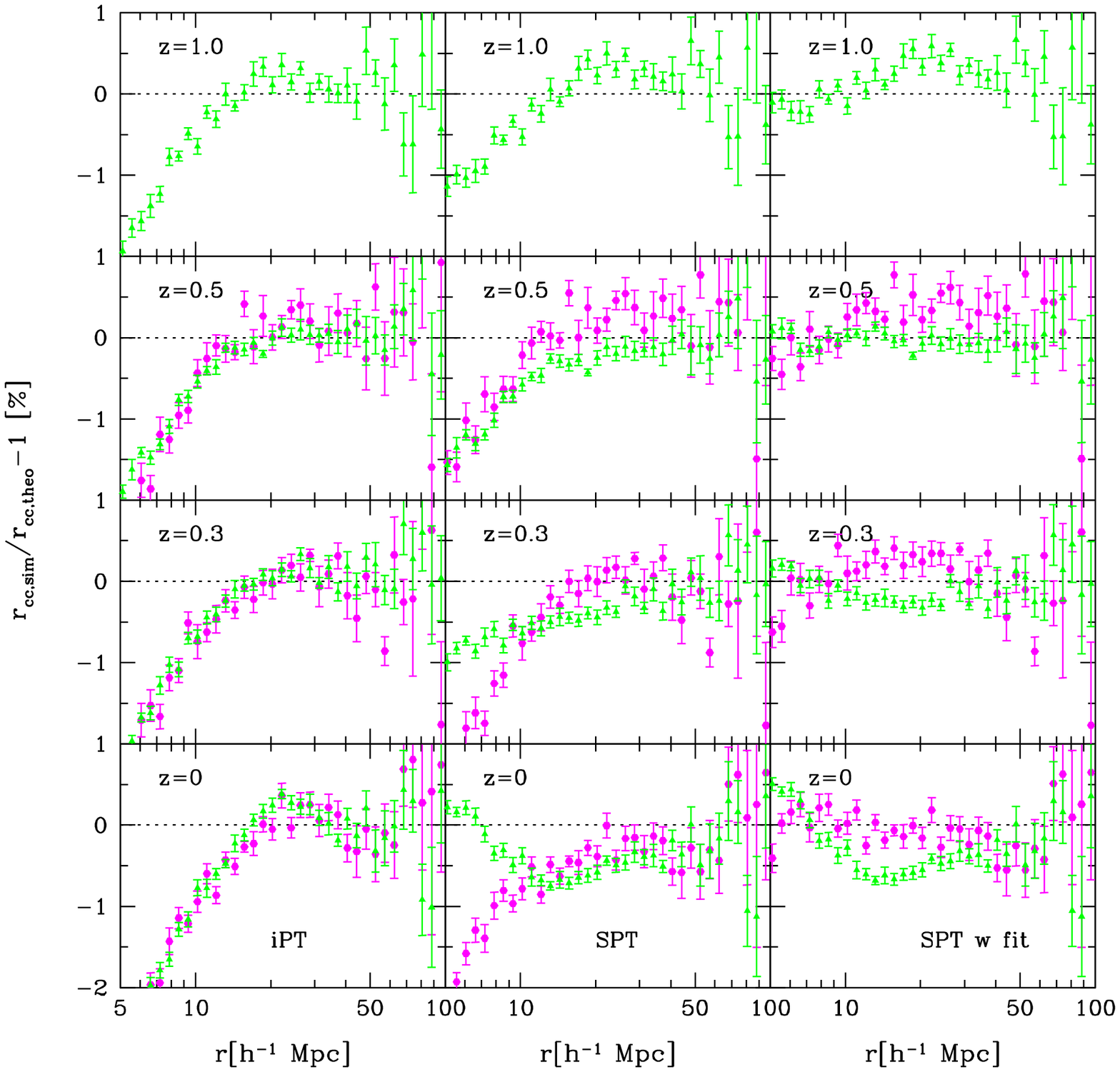}
\end{center}
\vskip-\lastskip
\caption{
Same as Figure~\ref{fig:cc.ratio}, but the results of Bin 4 (green
 triangle) and 5 (magenta circle) at
 $z=0$, 0.3, 0.5 and 1.0.
}
\label{fig:cc.ratioh}
\end{figure*}

In the framework of the local biasing model, the density field of galaxies and
their halos should be a {\it stochastic} function of the underlying
dark matter density field~\citep{1999ApJ...520...24D}. The
stochasticity is very weak on large scales, while it becomes more
important on small scales~\citep{1999ApJ...525..543M,
  2000ApJ...542..559T, 2001ApJ...558..520Y,2011MNRAS.412..995C}.

One of the characteristic parameters of stochasticity is the
cross-correlation coefficient between the matter and halo density
fields, defined as
\begin{equation}
 r_{\rm cc}(r)=\frac{\xi_{\rm hm}(r)}{\sqrt{\xi_{\rm mm}(r)\xi_{\rm hh}(r)}},
\label{cc}
\end{equation}
where $\xi_{\rm mm}(r)$, $\xi_{\rm hh}(r)$, and $\xi_{\rm hm}(r)$ are
the matter and halo auto-correlation functions, and their
cross-correlation function, respectively. The cross-correlation
coefficient is the
measure of the statistical coherence of the two
fields~\citep{1998ApJ...504..601P, 1999ApJ...518L..69T,
  1998ApJ...500L..79T, 2004MNRAS.355..129S,
  2009MNRAS.396.1610B,2012MNRAS.426..566C}. 
If any scale-dependent, deterministic, linear-bias model is assumed, we
have $r_{\rm cc}=1$. Therefore, deviations of the cross-correlation
coefficient from unity would arise due to both the nonlinearity and
stochasticity of bias.

Figure~\ref{fig:cc.zv3} shows the cross-correlation coefficient
between the matter and halo density fields at $z=0$, 0.3, 0.5, 1.0, 2.0 and
3.0. The cross, square, and triangle symbols are the $N$-body
simulation results measured from 30 realizations for halo masses
$1.37\le M_{\rm h}/(10^{12}{h}^{-1}M_{\odot})<4.11$ (Bin 1), $4.11\le
M_{\rm h}/(10^{12}{h}^{-1}M_{\odot})<12.32$ (Bin 2), and $1.23\le
M_{\rm h}/(10^{13}{h}^{-1}M_{\odot})<3.70$ (Bin 3), respectively.
The error bars describe the 1-$\sigma$ error on the mean value
obtained from 30 realizations. We do not plot the results in which the sum
of the 1-$\sigma$ error bars in a range of $5\le r_i/(h^{-1}{\rm
  Mpc})\le 100$ is larger than 0.12, i.e., $\sum_{5\le r_i\le
  100}\sigma_i>0.12$. It should be noted that halos in each bin are
more biased as redshift increases, because we impose the same halo
mass ranges for each bin. The solid curves show the iPT predictions.
The iPT obtains good agreements with simulation results down to
$r\sim$ 15$h^{-1}$Mpc within a range of error bars for all redshifts
and halo mass ranges we have considered. Particularly at $z=3.0$, the
iPT well reproduces the simulation result down to $r\sim$
6$h^{-1}$Mpc. 
The deviations from unity in the cross-correlation coefficient $r_{\rm cc}$ on
large scales are physical effects. Similar effects were also
predicted even in a simple model of local bias by \citet{1998ApJ...504..607S}. The iPT prediction for the deviations
has the same origin as theirs: the nonlinear dynamics on small scales
nontrivially affect the cross coefficients on very large scales. Our
simulations are consistent with these theoretical predictions.
Below, we will see fractional differences between
simulation results and theoretical predictions in
Figure~\ref{fig:cc.ratio}, to discuss the percentage error. The
difference between the iPT and simulation results on small scales
probably comes from the fact that the iPT breaks down on small scales
(see, Figure~\ref{fig:cor})~\citep{2011PhRvD..84d3501S,
  2013MNRAS.429.1674C}. One can see that the iPT prediction on small
scales is almost flat, unlike the simulation results. This is probably
because the asymptotic behaviors of the correlation functions based on
the iPT are almost the same (see Figure~\ref{fig:cor}),
and at any rate the iPT should not be applied on such small scales.

We also plot a simple model derived from Equations~(\ref{spthh}) and
(\ref{spthm}) as dotted curves and it is expressed
as~\citep{2010PhRvD..81f3531B}
\begin{equation}
 r_{\rm cc}(r)=1-\frac{1}{4}\left(\frac{b_2}{b_1}\right)^2\xi_{\rm L}(r),
\label{cc_bal}
\end{equation}
by using the approximations $A(r)\ll \xi_{\rm NL}(r)$ and $B(r)\ll \xi_{\rm NL}(r)$.

To empirically estimate $b_1$ and $b_2$, we use general relations between local bias parameters in Lagrangian
space and Eulerian space, which are derived in the spherical collapse model as~\citep{2011PhRvD..83h3518M}
\begin{align}
 b_1^{\rm E}&=b_1^{\rm L}+1,\label{bias_sph1}\\
 b_2^{\rm E}&=\frac{8}{21}b_1^{\rm L}+b_2^{\rm L},\label{bias_sph2}
\end{align}
where $b_1^{\rm E}$ and $b_2^{\rm E}$ are Eulerian bias parameters.
Note that both the Eulerian bias parameters $b_{n}^{\rm E}$ and the
Lagrangian bias parameters $b_{n}^{\rm L}$ are local and independent of scales.
In this phenomenological model, we simply substitute $b_1^{\rm E}$ and
$b_2^{\rm E}$ with $b_1$ and $b_2$. To calculate $b_1^{\rm L}$ and
$b_2^{\rm L}$, we use \citep{2008PhRvD..78h3519M}
\begin{equation}
  b_{n}^{\rm L} = 
  \frac{(-1)^n}{\delta_c^n}
  \frac{\displaystyle
    \int_{M_1}^{M_2}\nu^n\frac{\pd^{n}f_{\rm MF}(\nu)}{\pd\nu^n}
    \frac{\pd\ln\sigma(M)}{\pd{M}}
    \frac{\pd{M}}{M}}
  {\displaystyle
    \int_{M_1}^{M_2}f_{\rm MF}(\nu)
    \frac{\pd\ln\sigma(M)}{\pd{M}}\frac{\pd{M}}{M}}.
\label{b_func}
\end{equation}
for halos in a mass range $M_1\le M\le M_2$. The simple model
(Equation~\ref{cc_bal}) with the above estimates of bias parameters
shows better agreement with simulations for higher redshifts (i.e.,
more biased halos).
We find that the cross-correlation coefficients of halos with
$b_1^{\rm L} \gtrsim 2$ are well described in this method over all
scales we considered. For lower redshifts (i.e., less biased halos),
the simple model deviates more from the simulation results.



Meanwhile, when $b_1$ and $b_2$ are treated as free parameters, we fit
$b_2/b_1$ to the simulation results using a chi-square fit. The result
is shown as dotted lines in Figure~\ref{fig:cc.zfit}. Other lines and
symbols are the same as in Figure~\ref{fig:cc.zv3}. Fittings are done
in a range of $5\le r/(h^{-1}{\rm Mpc})\le 70$. In the fitting case,
an improvement from the above empirical method is little for cases of
high bias, but is important for cases of low bias. The simple model
with fitted bias replicates the simulation results over all scales at
$1\le z\le3$. We can see that the cross-correlation coefficients
estimated from $N$-body simulations have complicated behaviors in
quasilinear regimes at low redshifts, which cannot be described in
the simple model. We will describe percentage error later in
Figure~\ref{fig:cc.ratio}.

Figure~\ref{fig:cc.zfith} shows the results for the cross-correlation
coefficient of large halos with mass ranges $1.23\le M_{\rm
  h}/(10^{13}{h}^{-1}M_{\odot})<3.70$ (Bin 4) and $3.70\le M_{\rm
  h}/(10^{13}{h}^{-1}M_{\odot})<11.09$ (Bin 5) at redshift $z=0$, 0.3,
0.5, and 1.0. The triangle and circle symbols are the simulation
results of Bin 4 and Bin 5 estimated from 10 realizations of L2000.
The solid and dotted lines are the predictions of the iPT and SPT with
fitted bias, respectively.
As in Figures~\ref{fig:cc.zv3} and \ref{fig:cc.zfit}, the iPT shows
nice agreement with the simulation results on large scales even in
large halo masses. The simple model with fitting also reproduces the
simulation results for large halo masses.
However, the fitting values of $b_1/b_2$ are not, in general, the same as
those obtained from other statistics, such as the power spectrum and
bispectrum, because $b_1$ and $b_2$ are renormalized.

To clarify how well theoretical models predict the $N$-body results,
we plot fractional differences between $N$-body simulation results and
theoretical predictions, $[r_{\rm cc,sim}(r)-r_{\rm
    cc,theo}(r)]/r_{\rm cc,theo}(r)$, as shown in
Figures~\ref{fig:cc.ratio} and \ref{fig:cc.ratioh}. These figures show
that the iPT agrees with simulation results down to $r\sim$ 15
(10)$h^{-1}$Mpc within 0.5 (1.0) $\%$ for all redshifts and halo
masses we considered. It should be noted that the iPT does not have
any fitting parameter.
The SPT with empirically determined bias reproduces $N$-body
simulation results down to $r\sim$ 10$h^{-1}$Mpc within a
percent-level for all redshifts except for $z=0$ (see,
Figure~\ref{fig:cc.ratio}). In the SPT with bias determined by
fitting, a percent-level agreement is achieved over wide separation
angles for all redshifts. However, the fitted parameters $b_1$ and
$b_2$ are different from $b_1^{E}$ and $b_2^{E}$, which can be
determined by other methods, e.g., the power spectrum and bispectrum.

\section{Conclusion}
\label{sec:conc}
In this paper, we have used 40 large cosmological $N$-body simulations
of the standard $\Lambda$CDM cosmology to investigate the
cross-correlation coefficient between the halo and matter density fields
over a wide redshift range. The cross-correlation coefficient is
crucial to extract information of the matter density field by
combining galaxy clustering and galaxy-galaxy lensing measurements.
Since the first attempt to detect galaxy-galaxy lensing
\citep{1984ApJ...281L..59T}, its ability to constrain cosmological
parameters has been shown~\citep{2012arXiv1207.1120M}.

We compared the simulation results with theoretical predictions of the
iPT and simple models of bias with SPT. The iPT predicts the simulation
results down to $r\sim$ 15 (10)$h^{-1}$Mpc within 0.5 (1.0) $\%$ for
all redshifts and halo masses we considered. To improve the
prediction, the two-loop correction to the iPT might be important. In
the SPT with local bias model, bias parameters are renormalized and
therefore they are determined empirically or treated as free
parameters. The SPT with empirically determined biases with the spherical
collapse model shows better agreement with simulations for more biased
halos on small scales, although this model does not reproduce
the complicated behaviors of the simulation results on quasilinear scales
at low redshifts. The SPT with biases determined by fitting improves
the predictions but the situation is almost the same at low redshift.
Thus, the iPT accurately predicts the cross-correlation coefficient as
long as quasilinear scales are considered.

Let us finally comment on convolution Lagrangian perturbation theory
(CLPT), which was recently proposed by \citet{2013MNRAS.429.1674C}. The
CLPT applies additional resummations on top of the simple LRT
(restricted iPT with local Lagrangian bias), and its prediction
significantly improves the simple LRT for the correlation function in real
and redshift spaces on small scales. Therefore, it might be possible
that the CLPT gives a better prediction for the cross-correlation
coefficient between mass and halos and agrees with simulation results
on small scales. Although it is important to examine how well the CLPT
predicts these results, we leave it for future work.

In this paper we focused on fundamental features of bias stochasticity by
the methods of numerical simulations and theoretical models. We believe
the results of this paper could be a crucial first step to understand
the galaxy biasing for future precision cosmology.

\acknowledgments
We thank Uro\v s Seljak for useful comments.
M.S. is supported by a Grant-in-Aid for the Japan Society for Promotion of
Science (JSPS) fellows. T.M. acknowledges support from the
Ministry of Education, Culture, Sports, Science, and Technology (MEXT),
Grant-in-Aid for Scientific Research (C), No.~24540267, 2012. This
work is supported in part by a Grant-in-Aid for Nagoya University Global COE
Program, ``Quest for Fundamental Principles in the Universe: from
Particles to the Solar System and the Cosmos'', from the MEXT of
Japan.
Numerical computations were in part carried out on COSMOS provided by
 Kobayashi-Maskawa Institute for the Origin of Particles and the
 Universe, Nagoya University.

\bibliography{ms}


\end{document}